\documentclass[11pt]{article}
\usepackage[preprint]{acl}
\usepackage{times}
\usepackage[utf8]{inputenc}
\usepackage[T1]{fontenc}
\usepackage{microtype}
\usepackage{booktabs}
\usepackage{graphicx}
\usepackage{amsmath}
\usepackage{xcolor}
\usepackage{colortbl}
\definecolor{whours}{HTML}{FCEBD0}
\usepackage{algorithm}
\usepackage{algpseudocode}
\algrenewcommand\algorithmicrequire{\textbf{Input:}}

\title{When Sounds Turn Heads: Recovering Event Timing That Audio LLMs Fail to Say}
\author{Jongbhin Woo$^{*}$ \quad Jaewook Jeong$^{*}$ \quad Joon Son Chung \\
  KAIST \\
  \texttt{\{jongbin.woo,lockcept,joonson\}@kaist.ac.kr} \\
  {\normalfont\small $^{*}$Equal contribution.}}
\hypersetup{pdftitle={When Sounds Turn Heads: Recovering Event Timing That Audio LLMs Fail to Say},
  pdfauthor={Jongbhin Woo, Jaewook Jeong, Joon Son Chung}}

\begin{document}
\maketitle
\begin{abstract}
Audio LLMs can know when a queried sound occurs but fail to say it. We find a few attention heads that track the queried sound: they shift when the query changes, follow the sound when it moves, and removing or blocking them often hurts the model's timestamps. Yet the answer does not express what they track. Intervals read directly from these heads outperform the model's original answer and recover events that its timestamps miss. To bring this timing into the answer, we introduce head-guided decoding, which needs no training. It keeps the model's decision on whether the sound is present and rescores its candidate timestamps with a prior from these heads. Heads selected on development data transfer unchanged to new recordings and free-text queries, where head-guided decoding improves the timestamps of three audio LLMs, scores above their head readout on every benchmark, and, in most settings, outperforms random-head guidance. On recordings released after the models were trained, head-guided decoding raises mIoU from 0.07{}--0.36{} to 0.48{}--0.56{}. These results suggest that poor temporal answers in audio LLMs partly reflect a failure to say rather than to hear, and that the model's own attention can close part of this gap.
\end{abstract}

\section{Introduction}
\label{sec:intro}
\begin{figure}[!t]
\centering
\includegraphics[width=\columnwidth]{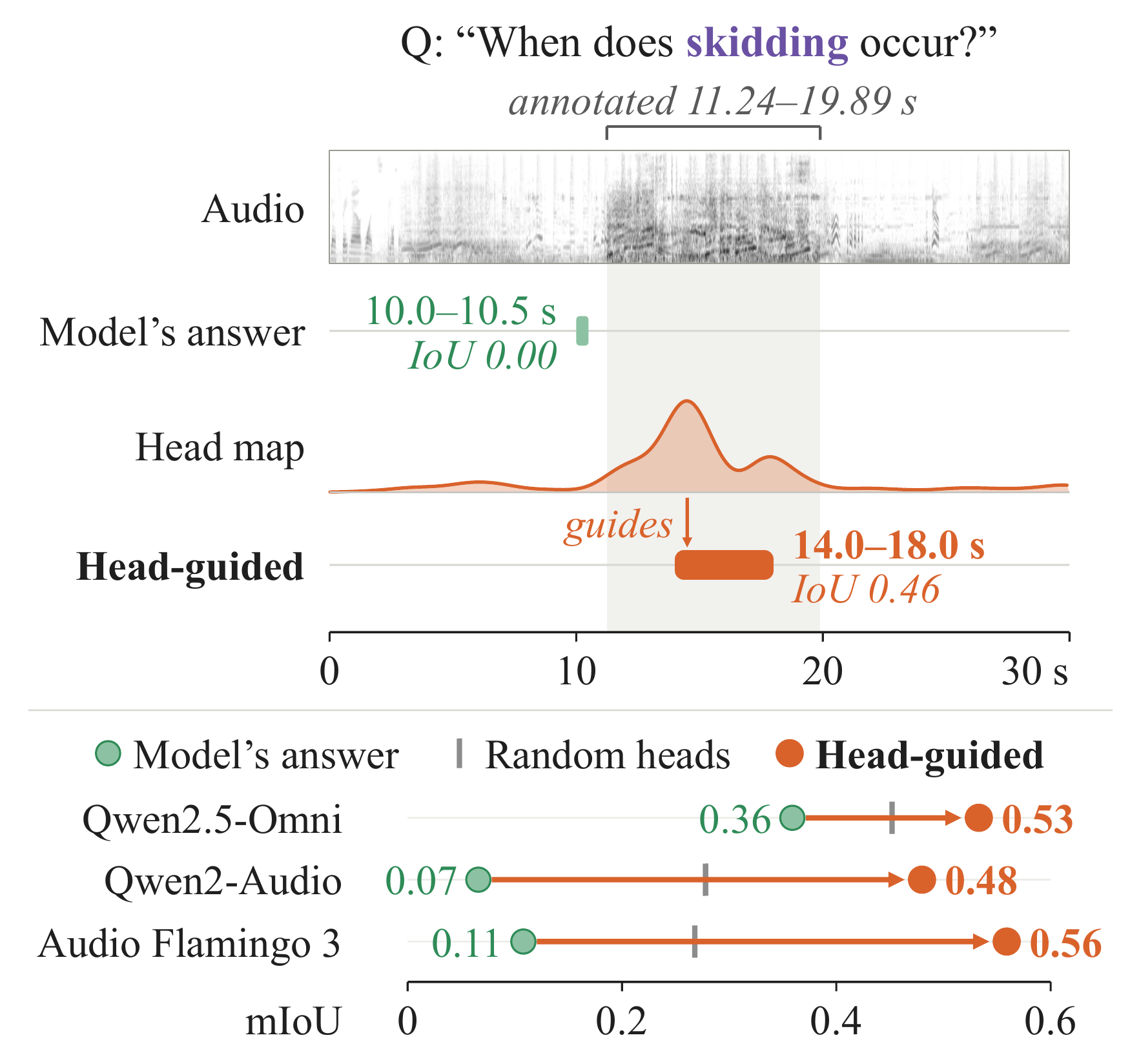}
\caption{\textbf{Heard but not said.} Top: on one recording, the attention of Qwen2.5-Omni's selected heads peaks inside the skidding sound, but the model's answer misses it. Head-guided decoding moves the answer inside. Bottom: on recordings released after the models were trained, head-guided decoding raises recording mIoU above the model's answer and random-head guidance (tick) for all three models, with settings tuned on UnAV-100 development data (Table~\ref{tab:main}).}
\label{fig:teaser}
\end{figure}

Given a recording and a sound query, temporal grounding identifies when the queried sound occurs. Knowing when a sound occurs supports audio search and indexing, home monitoring, and video editing and captioning, and recent work trains and evaluates audio LLMs on such questions \cite{timeaudio2025,spotsound2026,closer-look-failure-modes-temporal-understanding}.

Audio LLMs recognize sounds yet give biased timestamps and fail questions that compare their onsets, offsets, and durations \cite{sync-unveiling-temporal-bias-audio-chat,closer-look-failure-modes-temporal-understanding}. Asked when a skidding sound occurs, one model answers a time before the sound, while a few of its heads point inside the sound (Figure~\ref{fig:teaser}). Such an error can come from not hearing when the sound occurs or from not saying it. Failures to say appear beyond audio: language models can internally represent whether a statement is true and retain correct answers that they fail to express in their output \cite{burns2023discovering,azaria2023internal,orgad2025llms,gekhman2025insideout}, and multimodal models attend to the right image region, or hold visual information, that their answers miss \cite{mllms-know-where-look-training-free,hidden-plain-sight-vlms-overlook-their}. In video LLMs, a few attention heads locate the queried moment before the answer is generated \cite{mllms-know-when-before-speaking-revealing}. We investigate the gap between hearing and saying inside the model: we find attention heads that track the queried sound and test whether their timing reaches the answer.

We select attention heads whose attention to the audio moves to whichever sound the query names (query following, QF), and test whether they track the queried sound and contribute to the model's timestamp answers. First, their attention shifts toward the sound specified by the query text and follows the sound to its new location when it is moved. Second, blocking these heads' attention to the queried event weakens correct-interval preference more than blocking matched random heads. Finally, in the zeroing experiments, removing their outputs lowers generated-answer mIoU more than matched random-head removal in most comparisons. Read directly, these heads localize the queried sound more accurately than the model's original answer, so the timing is present but underused.

We therefore propose head-guided decoding, which preserves the model's decision on whether the sound is present and combines its candidate-answer likelihoods with a prior derived from the selected heads. Its design follows from three properties of temporal grounding in audio LLMs: heads attend even to silent input, the queried sound may be absent, and sounds are brief within long inputs. This prior scores agreement with the temporal map, allowing complete candidate answers to be rescored without updating model weights. For three audio LLMs, head-guided decoding scores above the head readout on every benchmark of Table~\ref{tab:main} and is significantly better than likelihood-only selection. In most settings, it is also significantly better than random-head guidance and published training-free baselines \cite{hsu2025reducingobjecthallucinationlarge,Leng_2024_CVPR,mllms-know-when-before-speaking-revealing}. For the question above, these results suggest that part of the error lies in saying rather than hearing, and that the model's own attention can close part of this gap.

We make three contributions:
\begin{itemize}
\item \textbf{A few attention heads track the queried sound.} Heads selected by query following move with the query and with the sound, and removing or blocking them hurts the model's timestamps in most comparisons (Section~\ref{sec:heads}).
\item \textbf{The answer does not express what these heads track.} Reading the heads alone localizes events better than the model's original answer and recovers events it misses on queries answered with a span (Section~\ref{sec:gap}).
\item \textbf{Head-guided decoding brings this timing into the answer without training.} Rescoring complete candidate answers with a head prior improves three audio LLMs over their original answers, scores above the head readout on every benchmark, and in most settings is significantly better than random-head guidance (Sections~\ref{sec:method}--\ref{sec:experiments}). On in-distribution and in-home recordings, the head prior alone carries this gain (Section~\ref{sec:gain-source}).
\end{itemize}

\section{Related Work}
\label{sec:related}
\paragraph{Temporal grounding in audio LLMs.}
Audio LLMs give biased timestamps and fail questions about the onsets, offsets, and durations of sounds \cite{sync-unveiling-temporal-bias-audio-chat,closer-look-failure-modes-temporal-understanding}, even when their later-layer attention aligns with the relevant speech \cite{sync-unveiling-temporal-bias-audio-chat}. Trained grounders add temporal encodings and markers \cite{timeaudio2025}, train for fine-grained temporal grounding \cite{spotsound2026}, learn frame-level modules on audio-LLM representations \cite{arxiv260915215}, or train frame-level heads that bypass timestamp tokens \cite{an2026encodeoncedecodenever}. Fine-tuning analyses separate event evidence from decoder accessibility \cite{semanticsreadout2026}. \emph{Steering Where to Listen} combines instruction-contrast residual steering with head-averaged later-layer attention, localizing on sequential synthetic clips with a sliding window of the true event duration \cite{steering-where-listen-instruction-based-activation}. We keep model weights unchanged, read unsteered per-head attention on natural recordings, and estimate both boundaries without the gold duration.

\paragraph{Attention-based localization across modalities.}
Multimodal LLMs attend to the image region a question needs even when their answers miss it \cite{mllms-know-where-look-training-free}, and in video, TG-Heads reads a few attention heads that locate the queried moment before the answer is generated \cite{mllms-know-when-before-speaking-revealing}. For Whisper \cite{radford2023robust}, word timestamps can be inferred by dynamic time warping over the cross-attention of selected decoder heads, which align the model's own transcript \cite{wagner2024crisperwhisper}. For audio, heads attend even to silence, the queried sound may be absent, and brief sounds sit within long inputs. We select query-following heads, test that they shape timestamps and, unlike these works, that they follow a moved sound, and design head-guided decoding around these properties, rescoring candidates without cropping or regenerating. A TG-Heads-style port is one of our baselines.

\paragraph{Decoding with internal signals.}
Decoding can use layer contrasts (DoLa) \cite{chuang2024doladecodingcontrastinglayers}, classifiers on attention ratios (Lookback Lens) \cite{chuang2024lookbacklensdetectingmitigating}, discriminator or expert guidance (FUDGE, DExperts) \cite{yang-klein-2021-fudge,liu-etal-2021-dexperts}, and input contrasts (VCD, CAD) \cite{Leng_2024_CVPR,shi-etal-2024-trusting}. For audio LLMs, audio-aware decoding (AAD) subtracts the prediction on silent input from the answer distribution \cite{hsu2025reducingobjecthallucinationlarge}. Head-guided decoding uses the same silent input differently: inside the model, to locate the sound, and adds this location evidence as a head prior that scores complete intervals, without additional learned parameters.

\paragraph{Functional heads and audio information flow.}
Sparse heads mediate induction, retrieval, and label anchoring \cite{context-learning-induction-heads,retrieval-head-mechanistically-explains-long-context,label-words-anchors-information-flow-perspective}, and specialized heads support translation \cite{voita-etal-2019-analyzing}. Audio studies identify specialist heads and use them to guide activation steering \cite{audio-language-models-listening-audio-specialist}, and trace information flow through fusion, conflict, and hubs \cite{causal-tracing-audio-text-fusion-large,who-wins-conflict-mechanistic-interpretability-text,probing-cross-modal-information-hubs-audio-visual-llms}. We focus on heads that track the queried event in time, testing event tracking with sound moves and contribution to timestamp answers with interventions. Attention alone need not explain predictions \cite{jain-wallace-2019-attention,wiegreffe-pinter-2019-attention}.

\section{Setup}
\label{sec:data}
\paragraph{Models.}
We study five{} audio LLMs: Qwen2.5-Omni (7B) [Omni] \cite{qwen2-omni-technical-report}, the instruction-tuned Qwen2-Audio (7B) \cite{qwen2audio2024}, Audio Flamingo 3 (8B) [AF3] \cite{audio-flamingo-advancing-audio-intelligence-fully}, Gemma-4 (E4B) [Gemma] \cite{gemma4team2026}, and Voxtral-Mini (3B) [Voxtral] \cite{liu2025voxtral}. Omni, Qwen2-Audio, and AF3 are our main models. Gemma and Voxtral test which findings extend beyond them (Table~\ref{tab:heads} and Section~\ref{sec:when-help}). Tables use full names. The model's answer is the timestamp the model generates on its own, or ``none''.\footnote{Code, selected heads, settings, and the evidence records behind every reported number will be publicly released.}

\paragraph{Data.}
Heads are selected once per model on a 23{}-recording subset of the UnAV-100 development set \cite{geng2023unav100} and reused unchanged on every dataset. We evaluate on two held-out UnAV-100 samples, on RealDESED in-home recordings released after the models were trained \cite{realdesed2026} (distinct from DESED \cite{turpault2019sound}), and on AudioGrounding free-text queries \cite{xu2021tag} (v1 and v2; all queries and short events).

\paragraph{Metrics.}
We score IoU between predicted and annotated interval unions, averaged per recording on UnAV-100 and RealDESED (recording mIoU) and per query on AudioGrounding all and short queries (query-mean union IoU). For queries whose target sound is present, predictions receive an IoU of zero if the model returns no interval or its answer cannot be parsed. We call a difference significant when its paired source-bootstrap 95\% interval excludes zero \cite{koehn-2004-statistical}. Intervals are not corrected for multiple comparisons.

\section{Attention Heads Track When Sounds Occur}
\label{sec:heads}
An audio LLM can misplace a sound in time for two reasons: the timing is missing inside the model, or it is present but not fully expressed in the answer. We look for it in attention heads, whose weights show where each token looks in the audio. We first identify heads whose attention follows the query (Section~\ref{sec:qf}), then test whether they follow the sound when it moves (Section~\ref{sec:tracking}), intervene on them to test their contribution to timestamps (Section~\ref{sec:causal}), and compare their readout with the model's answer (Section~\ref{sec:gap}).

\begin{figure*}[tp]
\centering
\includegraphics[width=0.977\textwidth]{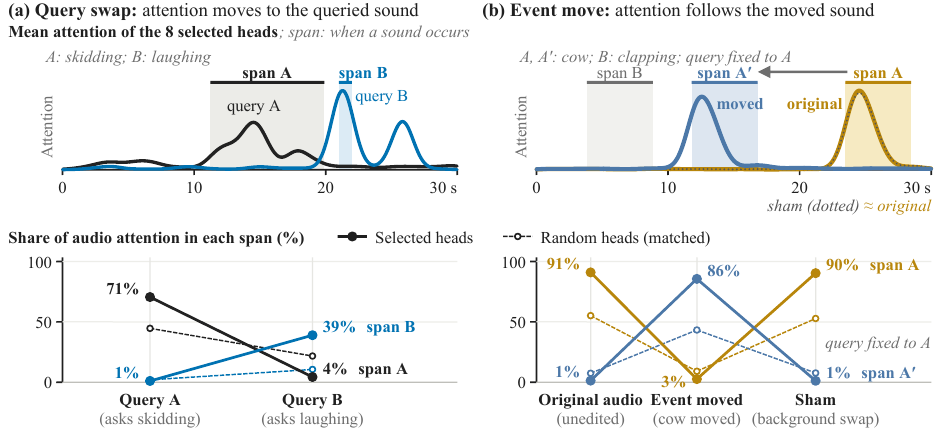}
\caption{\textbf{Selected heads follow the query and a moved sound} (Qwen2.5-Omni). Top: curves smoothed for display ($\sigma=0.8{}$~s). Bottom: unsmoothed shares; dashed: one matched random head set.}
\label{fig:qf}
\end{figure*}

\subsection{Query Swap: A Few Heads Follow the Query}
\label{sec:qf}
With the audio fixed, changing the target event in the text query shifts some heads' attention toward that event (Table~\ref{tab:heads}; Figure~\ref{fig:qf}(a)). We quantify this preference as query following.

\paragraph{Defining query following.}
For two distinct events occurring once in the same 30-s crop, let $I_A,I_B$ be annotated intervals and $q_A,q_B$ their queries. From the last token of the queried event label, we measure the fraction of head $h$'s attention to audio that falls in $I$, denoted $m_h(I\mid q)$. Audio bins crossing an interval boundary contribute in proportion to their overlap. We define

\begin{equation}
\begin{split}
\operatorname{QF}_h=\tfrac12\big[&m_h(I_A\mid q_A)-m_h(I_A\mid q_B)\\
+&m_h(I_B\mid q_B)-m_h(I_B\mid q_A)\big].
\end{split}
\label{eq:qf}
\end{equation}
Positive QF means that, on average, each event receives more attention when it is queried than when the other event is queried. We select the $K=8{}$ heads with highest mean QF on a head-selection set of 23{} recordings and call them the selected heads. Matched random head sets match head count, layer distribution, and audio attention measured on the head-selection set. Omni and AF3, which share the Qwen2.5-7B language model, share 6{} of their 8{} selected heads. Selected heads exceed the random QF reference on development recordings outside the head-selection set (Table~\ref{tab:heads}, \emph{Query swap QF}).

\begin{table}[!t]
\centering\fontsize{8}{9.6}\selectfont
\setlength{\tabcolsep}{1.5pt}
\begin{tabular}{@{}lcccccc@{}}
\toprule
& \multicolumn{2}{c}{\shortstack{Query swap\\QF}} & \multicolumn{2}{c}{\shortstack{Event move\\dest./origin}} & \multicolumn{2}{c}{\shortstack{Blocking\\$\Delta M$}} \\
\cmidrule(lr){2-3}\cmidrule(lr){4-5}\cmidrule(l){6-7}
Model & \cellcolor{whours}Sel. & Rand. & \cellcolor{whours}Sel. & Rand. & \cellcolor{whours}Sel. & Rand. \\
\midrule
 & \cellcolor{whours} &  & \cellcolor{whours}\textbf{.669} & .269 & \cellcolor{whours} &  \\
\raisebox{\dimexpr(\ht\strutbox+\dp\strutbox)/2\relax}[0pt][0pt]{Qwen2.5-Omni (7B)} & \cellcolor{whours}\raisebox{\dimexpr(\ht\strutbox+\dp\strutbox)/2\relax}[0pt][0pt]{\textbf{.207}} & \raisebox{\dimexpr(\ht\strutbox+\dp\strutbox)/2\relax}[0pt][0pt]{.072} & \cellcolor{whours}\textbf{.683} & .269 & \cellcolor{whours}\raisebox{\dimexpr(\ht\strutbox+\dp\strutbox)/2\relax}[0pt][0pt]{\textbf{.794}} & \raisebox{\dimexpr(\ht\strutbox+\dp\strutbox)/2\relax}[0pt][0pt]{.091} \\
\addlinespace[2pt]
 & \cellcolor{whours} &  & \cellcolor{whours}\textbf{.496} & .155 & \cellcolor{whours} &  \\
\raisebox{\dimexpr(\ht\strutbox+\dp\strutbox)/2\relax}[0pt][0pt]{Qwen2-Audio (7B)} & \cellcolor{whours}\raisebox{\dimexpr(\ht\strutbox+\dp\strutbox)/2\relax}[0pt][0pt]{\textbf{.150}} & \raisebox{\dimexpr(\ht\strutbox+\dp\strutbox)/2\relax}[0pt][0pt]{.026} & \cellcolor{whours}\textbf{.502} & .148 & \cellcolor{whours}\raisebox{\dimexpr(\ht\strutbox+\dp\strutbox)/2\relax}[0pt][0pt]{\textbf{.654}} & \raisebox{\dimexpr(\ht\strutbox+\dp\strutbox)/2\relax}[0pt][0pt]{$-$.075} \\
\addlinespace[2pt]
 & \cellcolor{whours} &  & \cellcolor{whours}\textbf{.558} & .171 & \cellcolor{whours} &  \\
\raisebox{\dimexpr(\ht\strutbox+\dp\strutbox)/2\relax}[0pt][0pt]{Audio Flamingo 3 (8B)} & \cellcolor{whours}\raisebox{\dimexpr(\ht\strutbox+\dp\strutbox)/2\relax}[0pt][0pt]{\textbf{.169}} & \raisebox{\dimexpr(\ht\strutbox+\dp\strutbox)/2\relax}[0pt][0pt]{.042} & \cellcolor{whours}\textbf{.586} & .189 & \cellcolor{whours}\raisebox{\dimexpr(\ht\strutbox+\dp\strutbox)/2\relax}[0pt][0pt]{\textbf{.480}} & \raisebox{\dimexpr(\ht\strutbox+\dp\strutbox)/2\relax}[0pt][0pt]{.064} \\
\addlinespace[2pt]
 & \cellcolor{whours} &  & \cellcolor{whours}\textbf{.140} & .052 & \cellcolor{whours} &  \\
\raisebox{\dimexpr(\ht\strutbox+\dp\strutbox)/2\relax}[0pt][0pt]{Gemma-4 (E4B)} & \cellcolor{whours}\raisebox{\dimexpr(\ht\strutbox+\dp\strutbox)/2\relax}[0pt][0pt]{\textbf{.037}} & \raisebox{\dimexpr(\ht\strutbox+\dp\strutbox)/2\relax}[0pt][0pt]{.014} & \cellcolor{whours}\textbf{.141} & .048 & \cellcolor{whours}\raisebox{\dimexpr(\ht\strutbox+\dp\strutbox)/2\relax}[0pt][0pt]{\textbf{.467}} & \raisebox{\dimexpr(\ht\strutbox+\dp\strutbox)/2\relax}[0pt][0pt]{$-$.264} \\
\addlinespace[2pt]
 & \cellcolor{whours} &  & \cellcolor{whours}\textbf{.310} & .079 & \cellcolor{whours} &  \\
\raisebox{\dimexpr(\ht\strutbox+\dp\strutbox)/2\relax}[0pt][0pt]{Voxtral-Mini (3B)} & \cellcolor{whours}\raisebox{\dimexpr(\ht\strutbox+\dp\strutbox)/2\relax}[0pt][0pt]{\textbf{.062}} & \raisebox{\dimexpr(\ht\strutbox+\dp\strutbox)/2\relax}[0pt][0pt]{.018} & \cellcolor{whours}\textbf{.315} & .080 & \cellcolor{whours}\raisebox{\dimexpr(\ht\strutbox+\dp\strutbox)/2\relax}[0pt][0pt]{\textbf{.090}} & \raisebox{\dimexpr(\ht\strutbox+\dp\strutbox)/2\relax}[0pt][0pt]{$-$.006} \\
\bottomrule
\end{tabular}

\caption{\textbf{Heads track events and shape interval preference.} Orange, bold: 8{} selected heads. Event move: destination (top) and origin (bottom). Rand.: matched references, not intervals (QF/destination/origin: p90 of 20; blocking: median of ten, nats).}
\label{tab:heads}
\end{table}

\subsection{Event Move: The Heads Follow the Sound}
\label{sec:tracking}
\textbf{Selected heads follow the queried sound when it moves in all five models.}
To test whether attention follows the sound itself, we move one event in a synthetic mixture and measure attention at its destination (Figure~\ref{fig:qf}(b)). The move--query interaction compares this gain against a background-only swap (sham) and against querying the other event. It exceeds the random reference (\emph{Event move}). Query-specific attention loss at the original location also exceeds the random p90 in all five models (Table~\ref{tab:heads}, \emph{Event move}, second line), supporting tracking through both destination gain and origin loss.

We also measure \emph{gold-span enrichment}: attention mass on the annotated span divided by the fraction of audio tokens it occupies. On UnAV-100 development recordings, including the head-selection set, selected-head enrichment is 1.59{}--1.66{} times that of matched random heads (median of ten sets) for Omni, Qwen2-Audio, and AF3. Section~\ref{sec:when-help} relates this ratio to when head-guided decoding helps.

\subsection{Interventions: The Heads Shape the Timestamp}
\label{sec:causal}
Tracking shows where the heads look, not that the timestamps depend on them. Selected heads contribute to timestamp answers in most comparisons: blocking their attention to the queried event weakens correct-interval preference, and zeroing their outputs lowers generated-answer mIoU.

\paragraph{Blocking event reads.}
We block selected heads' attention to the target event and measure the resulting decrease in correct-interval preference (the log-probability margin favoring the correct interval over another event's interval from the same recording). Table~\ref{tab:heads} shows a larger mean decrease for selected heads than for matched random heads in all five models.

\paragraph{Testing generated answers.}
We also set the selected heads' outputs to zero and let the model generate timestamps. Relative to layer-matched random-head removal, mIoU is lower in 10{} of 11{} main-model comparisons, with paired intervals below zero in 7{}.

\subsection{Head Readout: Knowing but Not Saying}
\label{sec:gap}\label{sec:readout-heads}
Heads that shape the timestamp may still hold more timing than the answer uses. Reading the main models' selected heads alone localizes events better than the generated timestamps and finds events they miss.

\begin{figure}[t]
\centering
\includegraphics[width=\columnwidth]{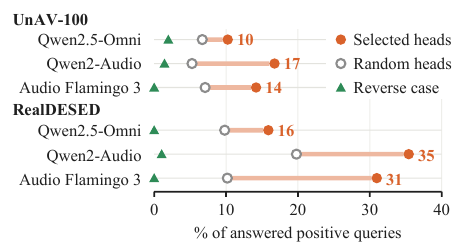}
\caption{\textbf{Knowing but not saying.} Share of answered positive queries on which the head readout is right (IoU $\ge$0.5{}) but the model's answer is wrong (IoU $<$0.1{}): filled circles for the selected heads, open circles for the median of ten matched random head sets. Triangles: the reverse case (answer right, head readout wrong) for the selected heads. AF3 UnAV-100: sample~2.}
\label{fig:readout-examples}
\end{figure}

\begin{figure*}[t]
\centering
\includegraphics[width=\textwidth]{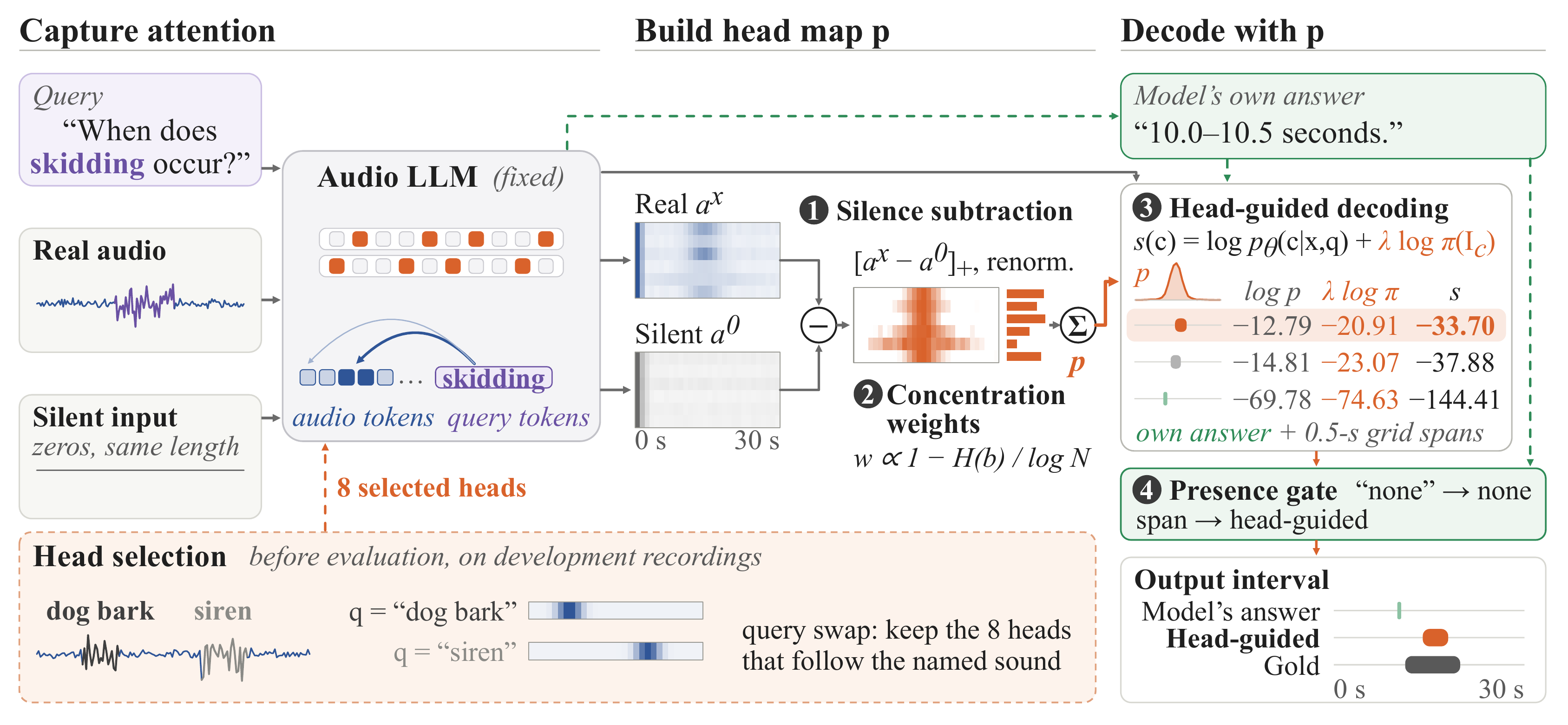}
\caption{\textbf{Overview of head-guided decoding.} Heads selected by query swap before evaluation give a head map $p$ from real-minus-silent attention (steps 1--2). Candidate answers are rescored by likelihood plus a head prior from $p$ (step 3), and the model's ``none'' is kept (step 4). Scores are measured. Maps, waveform, and heads are schematic.}
\label{fig:overview}
\end{figure*}

We form a temporal map $p$ over the $N$ audio bins. For each selected head $h\in\mathcal H$ and label-token row $r\in\mathcal R(q)$, let $a^x_{hr}$ and $a^0_{hr}$ be its attention on the real input $x$ and on a silent input $x^0$ of the same length, each renormalized over the audio bins after excluding the first two. We subtract silence, keep the positive part, renormalize, and average rows with concentration weights:
\begin{equation}
\begin{split}
b_{hr}&=\frac{[a^x_{hr}-a^0_{hr}]_+}{\lVert[a^x_{hr}-a^0_{hr}]_+\rVert_1},\\
p&=\textstyle\sum_{h,r}w_{hr}\,b_{hr},
\end{split}
\label{eq:map}
\end{equation}
where $w_{hr}\propto\max\big(0,\,1-H(b_{hr})/\log N\big)$ and $H$ is entropy, so concentrated rows count more.

\paragraph{The head readout scores above the answer.} We keep the model's span/none decision on whether the sound is present, which we call the presence gate, and replace its predicted boundaries with $\operatorname{decode}(p)$. This decoder smooths and thresholds the map, optionally expands the resulting regions, and returns the region with greatest unsmoothed mass, clipped to real audio. Its hyperparameters are selected on UnAV-100 development recordings. Pooled held-out recording mIoU rises from 0.210{} to 0.327{} for Omni, from 0.069{} to 0.213{} for Qwen2-Audio, and from 0.092{} to 0.311{} for AF3 (Table~\ref{tab:main}). Omni and Qwen2-Audio outperform 210{} matched random head sets on development data and UnAV-100 sample~1. All three models improve on RealDESED without refitting (Table~\ref{tab:main}, \emph{Head readout}).

\paragraph{The heads find events the answer misses.} Among positive queries that the model answers with a span, the head readout finds the event (IoU $\ge$0.5{}) while the answer misses it (IoU $<$0.1{}) on more RealDESED queries than matched random heads with their own readouts do, for every main model (Omni/Qwen2-Audio/AF3: 15.9{}/35.4{}/31.0{}\% versus 9.8{}/19.8{}/10.2{}\%; Figure~\ref{fig:readout-examples}). In all 6{} pooled UnAV-100 and RealDESED model--dataset cells the paired selected-minus-random interval lies above zero, and the reverse case, where the answer finds the event but the head readout misses it, stays at or below 2{}\%.

The heads therefore hold timing that the answer leaves unused.

\section{Head-Guided Decoding}
\label{sec:method}

The head readout selects among regions extracted by thresholding the head map $p$ (Section~\ref{sec:gap}). To compare boundaries beyond these regions, head-guided decoding instead scores candidate timestamp answers with the same map through a \emph{head prior}, which rewards map mass inside a candidate's intervals and penalizes their duration. It combines this prior with the model's likelihood of each candidate, using a guidance weight $\lambda$ selected on development data. Three properties of temporal grounding in audio LLMs shape its design. First, heads attend to audio positions even when no sound is present. Subtracting each head's attention on silent input, a content-free reference \cite{calibrate-before-use-improving-few-shot,found-middle-calibrating-positional-attention-bias}, removes this attention and keeps what the real audio adds (without it, Qwen2-Audio's RealDESED mIoU changes by $-0.121{}$; Table~\ref{tab:drivers}). Second, the queried sound may be absent, so decoding keeps the model's presence decision. Third, sounds are brief within long fixed-length inputs. Cropping around a sound would change the input and its time axis, and biasing attention during decoding stays near the model's answer (Table~\ref{tab:drivers}). We therefore rescore complete candidate intervals on the unchanged input. Figure~\ref{fig:overview} shows map construction (steps 1--2), candidate scoring (step 3), and the span decision (step 4).

\paragraph{Head-guided answer selection.}
Given audio $x$ and query $q$, let $c$ be a candidate answer and $I_c$ its parsed interval union. The model with parameters $\theta$ assigns probability $p_\theta(c\mid x,q)$ to the answer text. The head prior $\pi(I_c\mid x,q)$ assigns a positive compatibility score to its intervals, which need not be normalized. Their combined score is
\begin{equation}
s(c)=\log p_\theta(c\mid x,q)+\lambda\log\pi(I_c\mid x,q),
\label{eq:guided}
\end{equation}
where $\lambda$ controls the head contribution: $\lambda=0$ uses likelihood-only selection, and $\lambda=\infty$ denotes prior-only selection. The likelihood keeps answers that the model finds plausible, and the head prior pulls them toward where the selected heads attend. $\lambda$ sets this balance. The head prior carries the gain on UnAV-100 and RealDESED. On AudioGrounding, the likelihood adds to it (Section~\ref{sec:gain-source}). If the model's original answer contains no valid span (the presence gate), we return no interval. Otherwise we return the parsed intervals of $c^\star=\arg\max_{c\in\mathcal C(x,q)}s(c)$.
The candidate set $\mathcal C(x,q)$ consists of the model's original answer and all single intervals whose endpoints lie on a 0.5{}-s grid over the real audio of duration $S$, each at least 0.5{}~s long and written in the model's answer format. Algorithm~\ref{alg:guided} summarizes the procedure.

\begin{algorithm}[t]
\caption{Head-guided decoding}
\label{alg:guided}
\small
\begin{algorithmic}[1]
\Require audio $x$, query $q$, selected heads $\mathcal H$, settings $\lambda,\gamma,\epsilon$
\State $v\gets$ the model's answer to $(x,q)$
\If{$v$ contains no valid span}
\State \Return no interval \Comment{presence gate}
\EndIf
\State Capture the attention of $\mathcal H$ on $x$ and on silent $x^0$
\Statex \hspace{\algorithmicindent}and build the head map $p$ (Eq.~\ref{eq:map})
\State $\mathcal C\gets\{v\}\cup{}$single intervals on the candidate grid
\For{$c\in\mathcal C$} \Comment{Eqs.~\ref{eq:guided} and~\ref{eq:prior}}
\State $s(c)\gets\log p_\theta(c\mid x,q)+\lambda\log\pi(I_c)$
\EndFor
\Statex \hspace{\algorithmicindent}($\lambda=\infty$: rank by $\log\pi(I_c)$ alone)
\State \Return the parsed intervals of $\arg\max_{c\in\mathcal C}s(c)$
\end{algorithmic}
\end{algorithm}

\begin{table*}[t]
\centering\fontsize{7}{8.4}\selectfont
\setlength{\tabcolsep}{1pt}
\definecolor{whgain}{HTML}{1B5E20}%
\def\tnum#1{\makebox[13.9pt][r]{#1}}%
\def\tstack#1#2{\makebox[13.9pt][r]{\begin{tabular}[t]{@{}r@{}}#1\\[-1pt]\tgain{\textup{+}#2}\end{tabular}}}%
\def\tgain#1{{\fontsize{7}{7.7}\selectfont\color{whgain}#1}}%
\begin{tabular}{@{}lcccccccccccccccccc@{}}
\toprule
 & \multicolumn{6}{c}{\textbf{Qwen2.5-Omni (7B)}} & \multicolumn{6}{c}{\textbf{Qwen2-Audio (7B)}} & \multicolumn{6}{c}{\textbf{Audio Flamingo 3 (8B)}} \\
\cmidrule(lr){2-7}\cmidrule(lr){8-13}\cmidrule(l){14-19}
 & \multicolumn{1}{c}{In-dist.} & \multicolumn{5}{c}{Transfer} & \multicolumn{1}{c}{In-dist.} & \multicolumn{5}{c}{Transfer} & \multicolumn{1}{c}{In-dist.} & \multicolumn{5}{c}{Transfer} \\
\cmidrule(lr){2-2}\cmidrule(lr){3-7}\cmidrule(lr){8-8}\cmidrule(lr){9-13}\cmidrule(lr){14-14}\cmidrule(lr){15-19}
 & UnAV-100 & & \multicolumn{2}{c}{AG v1} & \multicolumn{2}{c}{AG v2} & UnAV-100 & & \multicolumn{2}{c}{AG v1} & \multicolumn{2}{c}{AG v2} & UnAV-100 & & \multicolumn{2}{c}{AG v1} & \multicolumn{2}{c}{AG v2} \\
\cmidrule(lr){4-5}\cmidrule(lr){6-7}\cmidrule(lr){10-11}\cmidrule(lr){12-13}\cmidrule(lr){16-17}\cmidrule(lr){18-19}
Method & (held-out) & RealDESED & All & Short & All & Short & (held-out) & RealDESED & All & Short & All & Short & (held-out) & RealDESED & All & Short & All & Short \\
\midrule
Model's answer & \tnum{.210} & \tnum{.359} & \tnum{.489} & \tnum{.295} & \tnum{.467} & \tnum{.337} & \tnum{.069} & \tnum{.066} & \tnum{.332} & \tnum{.141} & \tnum{.277} & \tnum{.151} & \tnum{.092} & \tnum{.108} & \tnum{.289} & \tnum{.151} & \tnum{.254} & \tnum{.165} \\
\midrule
\addlinespace[1pt]
\multicolumn{19}{@{}l}{\textbf{\textit{No head information}}} \\
Likelihood-only & \tnum{.207} & \tnum{.333} & \tnum{.524} & \tnum{.286} & \tnum{.494} & \tnum{.344} & \tnum{.093} & \tnum{.069} & \tnum{.370} & \tnum{.144} & \tnum{.341} & \tnum{.189} & \tnum{.156} & \tnum{.134} & \tnum{.315} & \tnum{.161} & \tnum{.315} & \tnum{.201} \\
AAD & \tnum{.253} & \tnum{.301} & \tnum{.535} & \tnum{.310} & \tnum{.522} & \tnum{.383} & \tnum{.084} & \tnum{.085} & \tnum{.379} & \tnum{.172} & \tnum{.336} & \tnum{.195} & \tnum{.182} & \tnum{.223} & \tnum{.397} & \tnum{.224} & \tnum{.374} & \tnum{.260} \\
VCD & \tnum{.245} & \tnum{.309} & \tnum{.529} & \tnum{.309} & \tnum{.524} & \tnum{\textbf{.387}} & \tnum{.067} & \tnum{.093} & \tnum{.364} & \tnum{.168} & \tnum{.330} & \tnum{.193} & \tnum{.151} & \tnum{.154} & \tnum{.401} & \tnum{.230} & \tnum{.382} & \tnum{.265} \\
\midrule
\addlinespace[1pt]
\multicolumn{19}{@{}l}{\textbf{\textit{Head attention}}} \\
Random heads & \tnum{.276} & \tnum{.452} & \tnum{.551} & \tnum{.300} & \tnum{.510} & \tnum{.356} & \tnum{.132} & \tnum{.278} & \tnum{\textbf{.477}} & \tnum{.219} & \tnum{.435} & \tnum{.266} & \tnum{.271} & \tnum{.268} & \tnum{.601} & \tnum{.354} & \tnum{.541} & \tnum{.392} \\
TG-Heads port & \tnum{.245} & \tnum{.346} & \tnum{.522} & \tnum{.314} & \tnum{.483} & \tnum{.351} & \tnum{.196} & \tnum{.292} & \tnum{.385} & \tnum{.261} & \tnum{.435} & \tnum{.326} & \tnum{.269} & \tnum{.335} & \tnum{.394} & \tnum{.275} & \tnum{.428} & \tnum{.342} \\
\rowcolor{whours}Head readout & \tnum{.327} & \tnum{.485} & \tnum{.536} & \tnum{.298} & \tnum{.492} & \tnum{.338} & \tnum{.213} & \tnum{.413} & \tnum{.457} & \tnum{.248} & \tnum{.463} & \tnum{.316} & \tnum{.311} & \tnum{.464} & \tnum{.538} & \tnum{.304} & \tnum{.504} & \tnum{.357} \\
\rowcolor{whours}Head-guided & \tstack{\textbf{.346}}{.136} & \tstack{\textbf{.533}}{.174} & \tstack{\textbf{.567}}{.078} & \tstack{\textbf{.325}}{.030} & \tstack{\textbf{.531}}{.063} & \tstack{.381}{.044} & \tstack{\textbf{.235}}{.166} & \tstack{\textbf{.480}}{.415} & \tstack{.461}{.130} & \tstack{\textbf{.286}}{.146} & \tstack{\textbf{.487}}{.210} & \tstack{\textbf{.368}}{.217} & \tstack{\textbf{.350}}{.258} & \tstack{\textbf{.559}}{.451} & \tstack{\textbf{.604}}{.315} & \tstack{\textbf{.388}}{.236} & \tstack{\textbf{.564}}{.309} & \tstack{\textbf{.434}}{.269} \\
\bottomrule
\end{tabular}

\caption{\textbf{Head-guided decoding improves every model's answer on every benchmark.} Recording mIoU; AudioGrounding (AG): query-mean union IoU. Comparators: Section~\ref{sec:setup}. Orange: ours; bold: best; small: gain over the answer.}
\label{tab:main}
\end{table*}
\paragraph{Head prior.}
We omit the conditioning on $x,q$ below. The head prior sums the map values inside $I$ to obtain $\mathcal{M}(I)$, so a candidate covering times with high $p_j$ scores higher. Since a longer candidate can collect more of the map simply by covering more audio, it subtracts a length penalty:
\begin{equation}
\log\pi(I)=\log(\epsilon+\mathcal{M}(I))-\gamma|I|/S.
\label{eq:prior}
\end{equation}
Here $|I|/S$ is the fraction of the real audio covered by the interval union, and $\gamma$ controls the penalty. The positive offset $\epsilon$ avoids taking the logarithm of zero. Because $p$ is normalized over the audio, the whole clip collects all of the map's mass. Without the penalty ($\gamma=0$), no shorter interval could score above it under the prior.

\section{Results}
\label{sec:experiments}
Head-guided decoding improves on every model's answer on every benchmark in Table~\ref{tab:main}, on RealDESED from 0.07{}--0.36{} to 0.48{}--0.56{} recording mIoU. It is significantly better than likelihood-only selection on every population, than random-head guidance in most settings, and than both contrastive decoding baselines in 15{} of the 18{} cells of Table~\ref{tab:main}. On held-out UnAV-100, the in-distribution test, it is significantly better than the model's answer, likelihood-only selection, and random-head guidance on all three models.
\subsection{Tuning and Comparators}
\label{sec:setup}
With each model's heads fixed, we select decoder and prior hyperparameters, including the guidance weight $\lambda$, on UnAV-100 development recordings. We evaluate the resulting configuration on held-out UnAV-100 recordings and transfer the hyperparameters unchanged to RealDESED. RealDESED inputs are 30-s windows, end-padded when a recording is shorter. Decoding uses the padded duration and every answer is clipped to the real audio before scoring. For AudioGrounding, we reselect these hyperparameters on its v1 training clips before evaluating on its v1 and v2 test sets.

We compare head-guided decoding with training-free comparators on the same audio LLM, with settings selected on the same data as ours: likelihood-only selection ($\lambda=0$); AAD \cite{hsu2025reducingobjecthallucinationlarge} and VCD \cite{Leng_2024_CVPR}, contrastive decoding against silent and noise-perturbed input; random-head guidance, the median over matched random head sets tuned separately; and the TG-Heads port, the map and interval procedure of TG-Heads \cite{mllms-know-when-before-speaking-revealing} on our selected heads (decode-only port: Table~\ref{tab:drivers}). We also report the post-hoc head readout as a reference. Every method keeps the model's span/none decision, so methods differ only in where they place a span.

\subsection{Transfer to In-Home Recordings}
\label{sec:transfer}
\textbf{Gains hold without retuning on in-home recordings released after the models were trained.} On RealDESED, head-guided decoding is significantly better than the model's answer, likelihood-only selection, and random-head guidance on all three models (Table~\ref{tab:main}). Random-head guidance already improves on the model's answer, so the difference from it is the gain specific to the selected heads. This head-specific gain ranges from $+0.081{}$ for Omni to $+0.291{}$ for AF3. Padding changes the model's answer far less than head-guided decoding does: the model's answer scores 0.376{}/0.063{}/0.143{} for Omni/Qwen2-Audio/AF3 at native length, against 0.359{}/0.066{}/0.108{} padded.

\subsection{Transfer to Free-Text Descriptions}
\label{sec:ag}
\textbf{Gains transfer to free-text descriptions.} Head-guided decoding is significantly better than likelihood-only selection for every model and AudioGrounding population, and than random-head guidance on short events of both versions and on v2 all queries for all three models. All methods share the 30-s padded input, which lowers AF3's v1 all-query answer from 0.495{} to 0.289{}. The exceptions against the model's answer and published baselines are on Omni, whose answer is already the strongest of the three.

\subsection{Where the Gain Comes From}
\label{sec:gain-source}
\begin{table}[t]
\centering\fontsize{8}{9.6}\selectfont
\setlength{\tabcolsep}{2pt}
\def\dnum#1{\makebox[2.3em][r]{#1}}%
\begin{tabular}{@{}lccc@{}}
\toprule
Method & {\fontsize{6.5}{7.8}\selectfont\shortstack{Qwen2.5-Omni\\(7B)}} & {\fontsize{6.5}{7.8}\selectfont\shortstack{Qwen2-Audio\\(7B)}} & {\fontsize{6.5}{7.8}\selectfont\shortstack{Audio\\Flamingo 3\\(8B)}} \\
\midrule
\rowcolor{whours}Head-guided (full, $K{=}8$) & \dnum{.533} & \dnum{.480} & \dnum{.559} \\
\midrule
\addlinespace[1pt]
\multicolumn{4}{@{}l}{\textbf{\textit{Which heads?}}} \\
\hspace{1pt}Labelled heads instead of QF & \dnum{+.019} & \dnum{.000} & \dnum{+.001} \\
\hspace{1pt}Single head ($K{=}1$) & \dnum{$-$.074} & \dnum{$-$.117} & \dnum{$-$.086} \\
\hspace{1pt}Four heads ($K{=}4$) & \dnum{+.017} & \dnum{$-$.038} & \dnum{+.008} \\
\hspace{1pt}Sixteen heads ($K{=}16$) & \dnum{+.012} & \dnum{$-$.025} & \dnum{+.014} \\
\hspace{1pt}Random heads (tuned) & \dnum{$-$.081} & \dnum{$-$.202} & \dnum{$-$.291} \\
\midrule
\addlinespace[1pt]
\multicolumn{4}{@{}l}{\textbf{\textit{How the map is built}}} \\
\hspace{1pt}No silence subtraction & \dnum{+.009} & \dnum{$-$.121} & \dnum{$-$.006} \\
\midrule
\addlinespace[1pt]
\multicolumn{4}{@{}l}{\textbf{\textit{How the heads are used}}} \\
\hspace{1pt}TG-Heads port (decode-only) & \dnum{$-$.173} & \dnum{$-$.413} & \dnum{$-$.427} \\
\midrule
\addlinespace[1pt]
\multicolumn{4}{@{}l}{\textbf{\textit{Reference}}} \\
\hspace{1pt}Model's answer & \dnum{.359} & \dnum{.066} & \dnum{.108} \\
\bottomrule
\end{tabular}

\caption{\textbf{What drives the gain.} RealDESED recording mIoU of full head-guided decoding and the model's answer; other rows: variant minus full. Labelled heads keep the selected settings. Orange: ours.}
\label{tab:drivers}
\end{table}
\textbf{Rescoring complete candidates, the core of head-guided decoding, drives the gain.} Table~\ref{tab:drivers} isolates the source of the gain on RealDESED: selecting heads with labels changes little, a single head scores lower for every model, and biasing attention during decoding (the decode-only TG-Heads port) stays near the model's answer. All 30{} matched random head sets score below the selected heads in development cross-validation.

On UnAV-100 and RealDESED the head prior carries the gain: prior-only selection, which ranks the same candidates by the head prior alone, is never significantly worse than the full method. On RealDESED, prior-only selection and head-guided decoding both score above the head readout for every model (e.g., Omni: 0.570{} and 0.533{} against 0.485{}). Finite-$\lambda$ rescoring adds a median of 12.35{}--22.20{} GPU-s per gate-open query on RealDESED, and prior-only selection ($\lambda=\infty$) needs no candidate likelihoods. On AudioGrounding, however, adding the model's likelihood yields significant gains over prior-only selection, by +0.022{} to +0.080{} mIoU, in all 9{} cells of all queries and the v1 single-interval subset. These gains suggest that, in this setting, the likelihood provides information beyond the attention mass and duration that the head prior captures.

\subsection{When the Heads Help}
\label{sec:when-help}
Gemma and Voxtral show when the selected heads help.

\textbf{The head readout scores above the answer in all five models.} In Gemma and Voxtral too, the head readout scores above the model's answer in every population (e.g., RealDESED: 0.013{} to 0.126{} for Voxtral, 0.044{} to 0.069{} for Gemma), as in every main-model cell of Table~\ref{tab:main}. What concentrates in the main models is the gain of head-guided decoding over random-head guidance.

\textbf{The main models' heads concentrate more on the event and gain more.} The main models' heads concentrate on the event (gold-span enrichment 1.59{}--1.66{} times matched random heads), while Gemma and Voxtral concentrate less (1.14{}--1.16{}), and their head-guided decoding is significantly better than random-head guidance on 3{} of 7{} and 0{} of 7{} populations. This association is descriptive. Presence detection does not follow this split: their balanced presence accuracy on UnAV-100 and RealDESED is close to AF3's. With oracle gates, their head readout stays near matched random heads on development data, but Voxtral's is significantly above them on RealDESED.

\section{Conclusion}
\label{sec:conclusion}
In all five audio LLMs we test, the head readout scores above the answer in every population, and in the main models their timestamp errors are partly failures to say rather than to hear. A few heads track the queried sound and often shape timestamps. On UnAV-100 and RealDESED, the main models' readout finds events the answer misses more often than matched random heads do, and head-guided decoding brings this evidence into the answer without training. The head-guided answer scores above the heads' own readout in every main-table cell. The main models' heads concentrate more on the event and gain more. We hope these findings motivate decoding methods that read what audio LLMs already hear.

\clearpage
\section*{Limitations}
\label{page:limitationsstart}
First, UnAV-100 and RealDESED queries are single-occurrence events of 1--10~s in 30-s crops, and AudioGrounding clips are short, so long recordings remain untested. Head-guided decoding also scores single-interval candidates, whose number grows quadratically with duration.
Second, on RealDESED and UnAV-100, gains over random-head guidance are large on the main models but small, absent or negative on Gemma and Voxtral (Section~\ref{sec:when-help}). The main models share Whisper-derived audio encoders at 40{}~ms per audio position, so with five{} models we cannot isolate the cause.
Third, head-guided decoding keeps the model's presence gate, so it cannot recover sounds the model reports as absent, and it needs attention access and labelled development data.
Finally, our interventions do not separate timing from event identity or answer production. When random heads remove a matched amount of attention, the blocking excess is not significant for Omni or for one of AF3's two random-head controls.

\section*{Ethical Considerations}
This work analyzes model attention and temporal predictions on existing sound-event datasets. It does not introduce a system for identifying people or inferring sensitive attributes. Better event timing could nevertheless support intrusive audio monitoring if deployed without consent. The observed grounding errors and annotation-absent responses make the method unsuitable as sole evidence in safety-critical or consequential decisions. Any release or reuse of audio should respect the source datasets' terms and the rights of recorded individuals. We report generalization limits rather than presenting internal attention as a guarantee of trustworthy detection.

\section*{Acknowledgments}
We used LLM assistants for code and writing. We reviewed the results and the manuscript and take full responsibility for the content.

\bibliography{references}
\end{document}